\documentclass[aps,preprint,preprintnumbers,amsmath,amssymb,amsfonts,groupedaddress,superscriptaddress,showpacs,showkeys,floatfix]{revtex4-1}

\usepackage{graphicx}
\usepackage{subfigure}
\usepackage{bm}% bold math
\usepackage{amsmath}
\usepackage{color}
\usepackage{array}
\usepackage{CJK}
\usepackage{ulem}

\newcommand{\bmath}{\begin{mathletters}}
\newcommand{\emath}{\end{mathletters}}
\newcommand{\be}{\begin{eqnarray}}
\newcommand{\ee}{\end{eqnarray}}
\newcommand{\ba}{\begin{array}}
\newcommand{\ea}{\end{array}}

\newcommand{\no}{\nonumber}

\newcommand{\calF} {\mathcal F}

\begin{document}
%\begin{CJK*}{GBK}{song}
\title{Experimental Determination of Electronic States via Digitized Shortcut-to-Adiabaticity and Sequential Digitized Adiabaticity}
%in a Superconducting Qubit Device"}

 \author{Ze Zhan}
   \thanks{These authors have contributed equally to this work.}
 \affiliation{Zhejiang Province Key Laboratory of Quantum Technology and Device, Department of Physics, Zhejiang University, Hangzhou, 310027, China}
   \author{Chongxin Run}
     \thanks{These authors have contributed equally to this work.}
 \affiliation{Zhejiang Province Key Laboratory of Quantum Technology and Device, Department of Physics, Zhejiang University, Hangzhou, 310027, China}
  \author{Zhiwen Zong}
  \affiliation{Zhejiang Province Key Laboratory of Quantum Technology and Device, Department of Physics, Zhejiang University, Hangzhou, 310027, China}
\author{Liang Xiang}
 \affiliation{Zhejiang Province Key Laboratory of Quantum Technology and Device, Department of Physics, Zhejiang University, Hangzhou, 310027, China}
   \author{Ying Fei}
 \affiliation{Zhejiang Province Key Laboratory of Quantum Technology and Device, Department of Physics, Zhejiang University, Hangzhou, 310027, China}
  \author{Zhenhai Sun}
  \affiliation{Zhejiang Province Key Laboratory of Quantum Technology and Device, Department of Physics, Zhejiang University, Hangzhou, 310027, China}
   \author{Yaozu Wu}
 \affiliation{Zhejiang Province Key Laboratory of Quantum Technology and Device, Department of Physics, Zhejiang University, Hangzhou, 310027, China}
 \author{Zhilong Jia}
 \affiliation{Key Laboratory of Quantum Information, University of Science and Technology of China, Hefei, 230026, China}
 \author{Peng Duan}
 \affiliation{Key Laboratory of Quantum Information, University of Science and Technology of China, Hefei, 230026, China}
 \author{Jianlan Wu }
 \email{jianlanwu@zju.edu.cn}
 \affiliation{Zhejiang Province Key Laboratory of Quantum Technology and Device, Department of Physics, Zhejiang University, Hangzhou, 310027, China}
 \author{Yi Yin}
 \email{yiyin@zju.edu.cn}
 \affiliation{Zhejiang Province Key Laboratory of Quantum Technology and Device, Department of Physics, Zhejiang University, Hangzhou, 310027, China}
 %\affiliation{Collaborative Innovation Center of Advanced Microstructures, Nanjing, 210093, China}
 \author{Guoping Guo}
 \email{gpguo@ustc.edu.cn}
 \affiliation{Key Laboratory of Quantum Information, University of Science and Technology of China, Hefei, 230026, China}
 \affiliation{Origin Quantum Computing, Hefei, 230026, China}

\begin{abstract}
A combination of the digitized shortcut-to-adiabaticity (STA) and the sequential digitized adiabaticity  is implemented in a superconducting quantum device
to determine electronic states in two example systems, the H$_2$ molecule and the topological Bernevig-Hughes-Zhang
(BHZ) model. For H$_2$, a short internuclear distance is chosen as a starting point,
at which the ground and excited states are obtained via the digitized STA.
From this starting point, a sequence of internuclear distances is built. The eigenstates at each distance
are sequentially determined from those at the previous distance via the digitized adiabaticity, leading to the potential energy landscapes
of H$_2$. The same approach is applied to the BHZ model,
and the valence and conduction bands are excellently obtained along the X-$\Gamma$-X linecut of the first Brillouin zone.
Furthermore, a numerical simulation of this method is performed to successfully extract the ground states of hydrogen chains
with the lengths of 3 to 6 atoms.

\end{abstract}

\maketitle

\section{Introduction}
\label{sec_01}
The rapid  advancement in quantum computation has shed the light on rich practical
applications~\cite{ChuangBook,GeorgescuRMP14,HouckNP12,GoogleNat19}. In a conventional computer, the
computational burden of an electronic state increases exponentially
with the system size. The evolution of a quantum state in a quantum computing
device can circumvent the difficulty of searching a multi-dimensional
Hilbert space~\cite{ChuangBook}. The electronic ground state can be determined
by the criterion of the lowest energy,  according to which
a variational quantum eigensolver (VQE) protocol has been proposed for a quantum
computing device~\cite{PeruzzoNC14,MalleyPRX16,KandalaNat17,GoogleSci20}. The VQE has been experimentally implemented
to calculate various systems ranging from small molecules (e.g., H$_2$~\cite{MalleyPRX16},
LiH and BeH$_2$~\cite{KandalaNat17}) to an atom chain of H$_{12}$~\cite{GoogleSci20}.
Due to the intrinsic requirement of energy minimization, it is more difficult to calculate
the electronic excited states with VQE. In a recent experiment, the excited eigenstates of H$_2$
are determined in a space composed of local states excited from the ground state~\cite{CollessPRX18}.

An important family of quantum computing methods is rooted in quantum adiabaticity~\cite{FarhiSci01,PengPRL08,YoungPRL10,VeisaJCP14,ArnabRMP08,JohnsonNature11,BoixoNC13,BarendsNat16,SteffenPRL03,LanyonSci11,SalathePRX15,HuNPJQI20}.
A quantum system evolves along its instantaneous eigenstate under a slowly varying Hamiltonian
and both dynamic and geometric phases are accumulated over time.
Alternatively, a shortcut-to-adiabaciticy (STA) is designed to accelerate the adiabatic process
with the assistance of a counter-diabatic Hamiltonian~\cite{BerryJPhysA09,XChenPRL2010,OdelinRMP19,TakahashiPRA17,ZhangSciRep16,ZZXPRA17,SCPMA2018,WangNJP2018,ZZXNJP2018,PWPRL19,
chenxiPRAPP20,SantosJPAMT18,SmithNJP18,chenxiPRL10}.
In superconducting quantum devices, the experimental implementation of the STA  has been well studied in a single-qubit system~\cite{ZhangSciRep16,ZZXPRA17,SCPMA2018,WangNJP2018,ZZXNJP2018}
but becomes difficult in multi-qubit systems where the exact form of the counter-diabatic Hamiltonian is often complicated.
Recently, a single-qubit approximation~\cite{chenxiPRAPP20} and a sum of nested commutators~\cite{PWPRL19} have been proposed
to simplify the estimation of the counter-diabatic Hamiltonian.
The realization of a time-varying multi-qubit Hamiltonian is however an experimentally nontrivial task.
On the other hand, digital quantum computing
based on the Trotter-Suzuki decomposition offers a universal tool for various quantum
problems~\cite{BarendsNC15,LloydSci96,TrotterPAMS59,SuzukiCMP76}.
With continual improvement in equipments and algorithms, digital quantum computing has gradually
become practically available. For example, the digitized adiabatic~\cite{BarendsNat16,SteffenPRL03,LanyonSci11,SalathePRX15}
and STA~\cite{chenxiPRAPP20} algorithms have been proposed and tested in the preparation of entangled multi-qubit states.

To fully understand the electronic structure of a quantum system, the eigenstates and eigenenergies need to
be determined over a whole parameter space. A potential energy landscape of a microscopic system can lead to its molecular
structure or relevant reaction pathways. The electronic band structure of a crystal system is essential for
electromagnetic properties. With the development of various eigensolvers, we may
determine the eigenstates repeatedly via the same method as the parameter changes.
Instead, the historical information collected over the eigensolving process can be utilized for a later state determination.
In particular, we introduce a method of sequential digitized adiabaticity. For a pre-selected parameter sequence,
the eigenstates at each position can be determined by a digitized adiabatic evolution from the eigenstates predetermined  at the
previous position. Due to a short distance between two adjacent positions, the adiabatic process can be sufficiently reliable
over a short operation time.  In this paper, we apply the combination of the digitized STA and the sequential digitized adiabaticity
to experimentally determine the potential energy landscapes (both the ground and first excited states) of H$_2$
and the valence and conduction bands along the X-$\Gamma$-X linecut for the topological Bernevig-Hughes-Zhang (BHZ) model~\cite{ZhangSci06}.
Furthermore, we explore the ground state energy landscapes of multi-atom hydrogen chains by a numerical simulation.

\section{Theory}
\label{sec_02}

\begin{figure}[tp]
\centering
 \includegraphics[width=0.7\columnwidth]{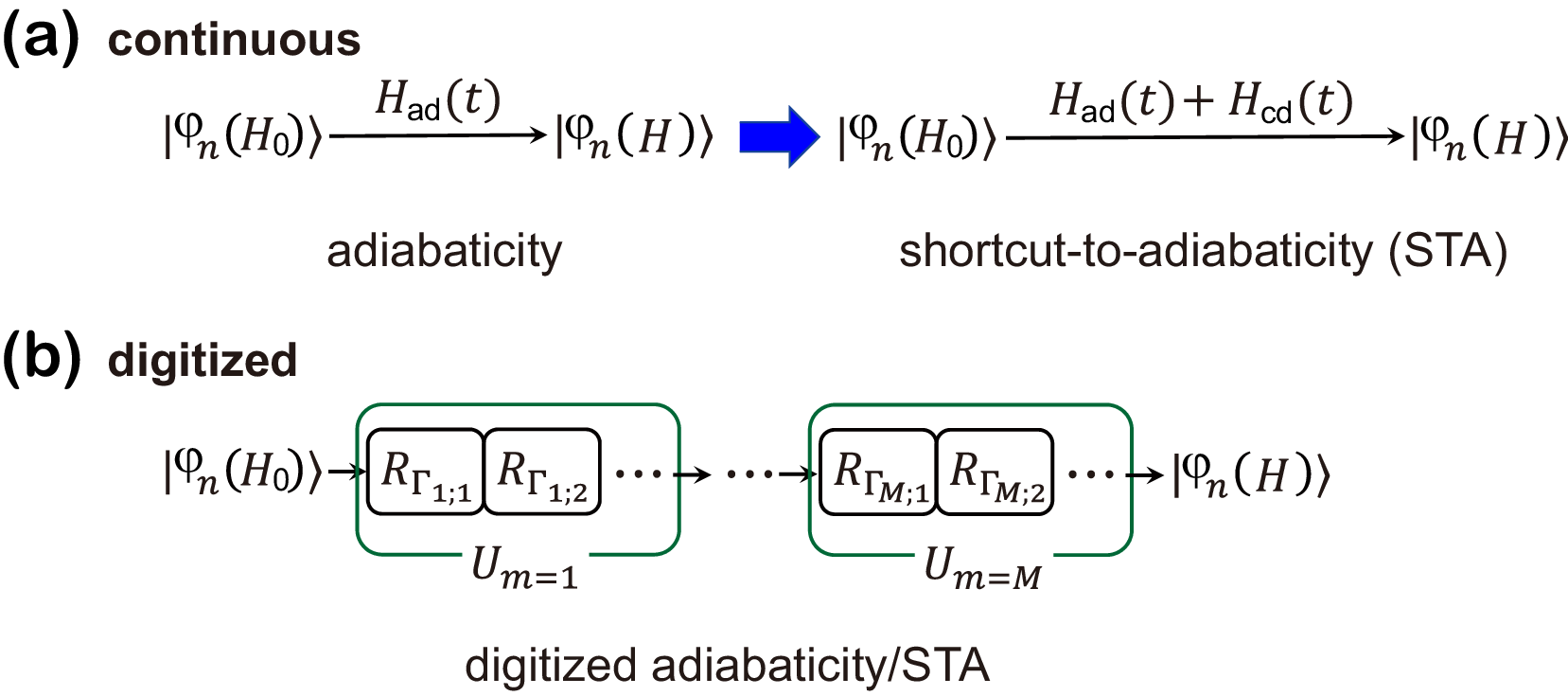}
\caption{A schematic diagram of (a) continuous and (b) digitized adiabatic/STA processes.}
\label{fig_01}
\end{figure}

A digitized adiabatic or STA eigensolver is rooted in an adiabatic process (see Fig.~\ref{fig_01}).
To extract the eigenenergies and eigenstates $\{\varepsilon_n, |\varphi_n\rangle\}$ of
a target Hamiltonian $H$, we begin with an initial Hamiltonian $H_0$ and design a time-varying Hamiltonian,
\be
H_\mathrm{ad}(t)=H_0+\lambda(t)(H-H_0),
\label{eq_00}
\ee
where the coefficient satisfies $\lambda(t\!=\!0)\!=\!0$ at the initial time and $\lambda(t\!=\!T)\!=\!1$ at the final time.
If the initial state is an eigenstate, $|\Psi(t=0)\rangle=|\varphi_n(H_0)\rangle$, and the Hamiltonian varies extremely slowly, $|\dot{\lambda}(t)|\rightarrow 0$, the adiabatic evolution spontaneously drags the quantum system to its final eigenstate, %$|\Psi(t\!=\!T)\rangle\!\propto\!|\varphi_n(H)\rangle$.
$\lim_{T\rightarrow\infty}|\Psi(t\!=\!T)\rangle\!=\!U_\mathrm{ad}|\Psi(0)\rangle\!=\!\exp[i\Phi_n(T)]|\varphi_n(H)\rangle$.
Here $U_\mathrm{ad}=\exp_+[-i\int_0^T H_\mathrm{ad}(\tau)d\tau]$ is the time evolution operator and
the phase $\Phi_n(T)$ includes both dynamic and geometric parts.
The reduced Planck constant $\hbar$ is set to be unity throughout this paper.

For the STA process, a counter-diabatic Hamiltonian is introduced as~\cite{BerryJPhysA09,XChenPRL2010,OdelinRMP19,TakahashiPRA17,ZhangSciRep16,ZZXPRA17,SCPMA2018,WangNJP2018,ZZXNJP2018,PWPRL19,
chenxiPRAPP20,SantosJPAMT18,SmithNJP18,chenxiPRL10}
\be
H_\mathrm{cd}(t)=i\sum_n \left[|\dot{\varphi}_n(t)\rangle\langle \varphi_n(t)|-\langle \varphi_n(t)|\dot{\varphi}_n(t)\rangle |\varphi_n(t)\rangle\langle \varphi_n(t)|  \right]
\label{eq_01}
\ee
where $|\varphi_n(t)\rangle\equiv|\varphi_n(H_\mathrm{ad}(t))\rangle$ is the $n$-th instantaneous eigenstate of the adiabatic Hamiltonian $H_\mathrm{ad}(t)$.
For a finite operation time $T$, the quantum system subject to the total Hamiltonian $H_\mathrm{tot}(t)\!=\!H_\mathrm{ad}(t)+H_\mathrm{cd}(t)$
is dragged exactly to the eigenstate $|\Psi(t\!=\!T)\rangle\!=\!U_\mathrm{STA}|\Psi(0)\rangle\!=\!\exp[i\Phi^\prime_n(T)]|\varphi_n(H)\rangle$
under the conditions of $\dot{\lambda}(0)\!=\!0$ and $\dot{\lambda}(T)\!=\!0$. The time evolution operator is changed to be
$U_\mathrm{STA}=\exp_+[-i\int_0^T H_\mathrm{tot}(\tau)d\tau]$.
Although the counter-diabatic Hamiltonian has been derived under various frameworks, e.g., the elimination of nonadiabatic transition~\cite{BerryJPhysA09,XChenPRL2010,OdelinRMP19,TakahashiPRA17,ZhangSciRep16,ZZXPRA17,SCPMA2018,WangNJP2018,ZZXNJP2018,PWPRL19,
chenxiPRAPP20,SantosJPAMT18,SmithNJP18}
and the Lewis-Riesenfeld invariant~\cite{chenxiPRL10}, $H_\mathrm{cd}(t)$ can recover the same expression in Eq.~(\ref{eq_01}).

From the practical perspective, the adiabatic process usually needs a long operation time,
i.e. $\varepsilon T_\mathrm{min}\gg 1$ with $\varepsilon$ a characteristic
energy of the system and $T_\mathrm{min}$ the minimum value of $T$ for the output fidelity above a threshold.
The STA can accelerate the adiabatic process but the appearance of instantaneous eigenstates in Eq.~(\ref{eq_01})
contradicts our ultimate goal of determining $|\varphi_n(H)\rangle$. To avoid this problem,
we apply a single-qubit assumption in the scenario of a weak inter-qubit interaction
(see  Appendix~\ref{appA} and Sec.~\ref{sec_03} for details)~\cite{chenxiPRAPP20}.
The operation time of this approximate STA can be compressed to satisfy $\varepsilon T_\mathrm{min}\lesssim O(1)$.
With the increase of the inter-qubit interaction, a more systematic treatment
is an expansion over nested commutators~\cite{PWPRL19}, which requires a heavy computational cost as
the expansion order increases. Instead, we re-select a more appropriate initial Hamiltonian $H_0$
to suppress the Hamiltonian deviation $\|H-H_0\|$ and the state distance between
$|\varphi_n(H)\rangle$ and $|\varphi_n(H_0)\rangle$. The adiabatic  prediction is practically reliable
under a short operation time [$\varepsilon T_\mathrm{min}\lesssim O(1)$] without the necessity of the STA.

The experimental realization of a time-varying multi-qubit Hamiltonian (adiabatic or STA) is possible but difficult.
Instead, a digitized algorithm can realize the time evolution of an arbitrary Hamiltonian~\cite{BarendsNat16,SteffenPRL03,LanyonSci11,SalathePRX15,chenxiPRAPP20}.
After dividing the operation time into $M$ segments ($\Delta t\!=\!T/M$),
we apply the Trotter-Suzuki decomposition~\cite{BarendsNC15,LloydSci96,TrotterPAMS59,SuzukiCMP76} to the time evolution operator ($U=U_\mathrm{ad}$ or $U_\mathrm{STA}$),  %$U\!=\!T_+\exp[-i\int_0^T H_\mathrm{tot}(t)dt]$,
which leads to
\be
U\!\approx\!U_M\cdots U_2 U_1
\label{eq_02}
\ee
with $U_{m=1,\cdots,M}\!=\!\exp[-i H_\mathrm{ad/tot}(m\Delta t)\Delta t]$.
In general, the adiabatic or total Hamiltonian can be expanded into $H_\mathrm{ad/tot}(t)\!=\!\sum_{j=1}^{J} \omega_j(t) \Gamma_j/2$,
where $\Gamma_j$ is a quantum operator, $\omega_j(t)$ is its associated coefficient and $J$ is the total number of operators.
The Trotter-Suzuki decomposition allows a new factorization~\cite{BarendsNC15,LloydSci96,TrotterPAMS59,SuzukiCMP76},
\be
U_m\!\approx\!R_{\Gamma_J}(\theta_{m; J})\cdots R_{\Gamma_2}(\theta_{m; 2}) R_{\Gamma_1}(\theta_{m; 1})
\label{eq_03}
\ee
where $R_{\Gamma_j}(\theta_{m; j})\!=\!\exp[-i\theta_{m; j}\Gamma_j/2]$ can be viewed as a generalized rotation operator
over an angle $\theta_{m; j}\!=\!\omega_j(m\Delta t)\Delta t$.
In practice, $R_{\Gamma_j}(\theta_{m; j})$ can be realized by a combination of sequential single- and two-qubit gates.
Notice that the real experimental time is determined by the total time of gates rather than the adiabatic or STA operation time $T$.
%{\color{blue}%The minimum number of steps $M$ for convergence is positively dependent on the necessary
%operation time $\varepsilon T_\mathrm{min}$ of the continuous adiabatic or STA process.
%However, the Trotter-Suzuki decomposition induces a fluctuating behavior on the output fidelity
%with the change of the operation time $T$. Consequently, the value of $T$ needs to be optimized
%for the best performance for a fixed $M$.
%}

\section{EXPERIMENTAL realization}
\label{sec_03}

The digitized adiabatic and STA eigensolvers are implemented in a quantum computing device
composed of six superconducting cross-shaped transmon qubits~\cite{WangPRAPP2019},
from which two qubits ($Q_A$ and $Q_B$) are selected.
The ground and excited states of each qubit are mapped onto the up and down states of a spin, i.e.,
$|0\rangle\!\leftrightarrow\!|+\rangle$ and $|1\rangle\!\leftrightarrow\!|-\rangle$.
The device is mounted in an aluminum sample box and in a dilution refrigerator
with a base temperature around 10 mK. The frequency-tunable qubits are initially biased at an operation point
through two $Z$-control lines. In our experiment, the two qubits are set at $\omega_A/2\pi\!=\!6.21$ GHz
and $\omega_B/2\pi\!=\!5.70$ GHz. The two anharmonicity parameters are $\Delta_{A}/2\pi\!\approx\!\Delta_B/2\pi\!=\!-250$ MHz,
enabling a selective microwave drive to control each qubit through its $XY$-control line. At the operation point,
the relaxation times are $T_{A;1}\!=\!5.8$ $\mu$s  and $T_{B;1}\!=\!6.9$ $\mu$s
while the pure dephasing times are $T_{A;\phi}\!=\!26$ $\mu$s and $T_{B;\phi}\!=\!28$ $\mu$s.
The readout fidelities of the ground and the excited states are $F_{A;0}\!=\!99\%$
and $F_{A;1}\!=\!95\%$ for qubit $A$, and $F_{B;0}\!=\!97\%$ and $F_{B;1}\!=\!93\%$ for qubit $B$~\cite{XLPRAPP2020}.

\subsection{Hydrogen Molecule}
\label{sec_031}

Under the Born-Oppenheim approximation, the electronic Hamiltonian of a hydrogen molecule is written as
\be
H(\vec{r}_1, \vec{r}_2|\vec{R}_1, \vec{R}_2) &=& \mathcal E_{\mathrm{nucl}}(R)+
\sum_{i=1,2}\frac{\vec{p}^2_i}{2m}-\sum_{i,j=1,2}\frac{e^2}{4\pi\varepsilon_0 |\vec{r}_i-\vec{R}_j|}+\frac{e^2}{4\pi\varepsilon_0 |\vec{r}_1-\vec{r}_2|},
\label{eq_04}
\ee
where $\vec{p}_{i=1, 2}$ and $\vec{r}_{i=1,2}$ are the momentums and coordinates of two electrons, and $\vec{R}_{i=1,2}$ are the coordinates of two nuclei.
The nuclear term $\mathcal E_{\mathrm{nucl}}(R)$ is dependent on the internuclear distance $R=|\vec{R}_1-\vec{R}_2|$.
For the ground and several lowest excited electronic states of H$_2$, the single-electron basis functions are
combined by two delocalized orbitals $\{|\phi_g\rangle, |\phi_u\rangle\}$ and two spin states $\{|+\rangle, |-\rangle\}$,
i.e., $|\chi_{i=1,\cdots, 4}(\vec{r}, \sigma)\rangle\!=\!|\phi_{g/u}(\vec{r})\rangle\otimes|\sigma\!=\!+/-\rangle$
with $\vec{r}$ and $\sigma$ the coordinate and spin variables~\cite{PeruzzoNC14,MalleyPRX16,CollessPRX18}.
In the number representation with regard to $\{|\chi_{i}\rangle\}$, the second quantization of the Hamiltonian is
$H=\mathcal E_{\mathrm{nucl}}+\sum h_{ij}a^+_i a_j +\sum h_{ijkl} a^+_i a^+_j a_k a_l$,
where $a_i$ and $a^+_i$ are annihilation and creation operators satisfying
$[a_i, a^+_j]_+=\delta_{i, j}$. The single- and two-electron integrals ($h_{ij}$ and $h_{ijkl}$) depend on the
internuclear distance $R$~\cite{PeruzzoNC14,MalleyPRX16,CollessPRX18}.
%{\color{red}Following the second quantization, a Hamiltonian of fermions is re-expressed by interactions among qubits.}
As shown in Appendix~\ref{appB}, the Bravyi-Kitaev transformation~\cite{SeeleyJCP01} is applied and the target Hamiltonian is approximated as
\be
H =  g_0  + g Z_A + g Z_B + g_{12} Y_AY_B,
\label{eq_05}
\ee
where $\{X_\alpha, Y_\alpha, Z_\alpha\}$ is the set of Pauli matrices ($\alpha=A, B$),
and the parameters $\{g_0, g, g_{12}\}$ are functions of $R$~\cite{PeruzzoNC14,MalleyPRX16,CollessPRX18}.

\begin{figure}[tp]
\centering
 \includegraphics[width=0.7\columnwidth]{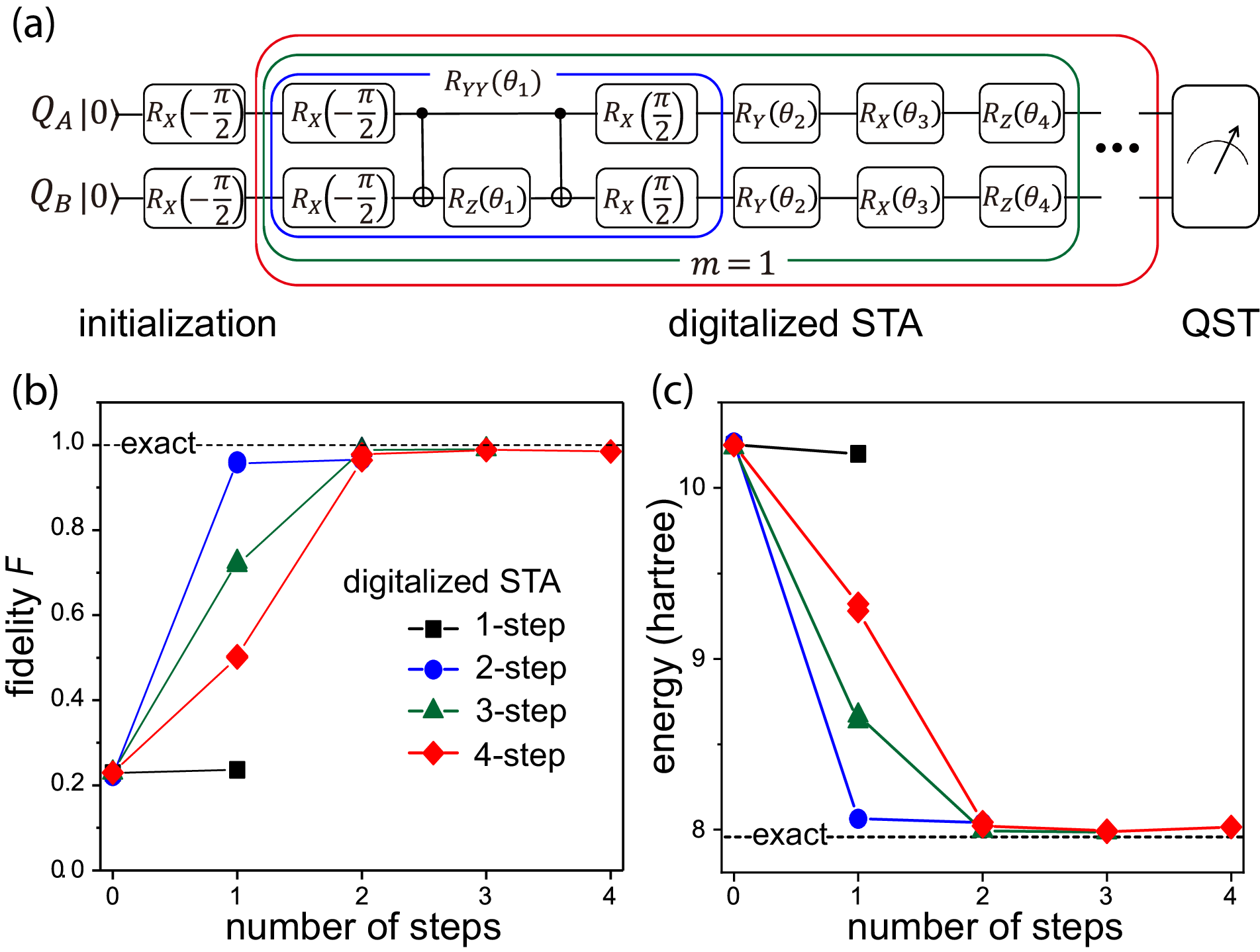}
\caption{Determination of the ground state of H$_2$ via a digitized STA eigensolver for the internuclear distance at $R=0.05$ \AA.
(a) A quantum circuit diagram of the $M$-step digitized STA process.
(b) The evolution of the state fidelities (in comparison to the exact ground state) in four digitized STA experiments ($1\le M\le 4$).
(c) The evolution of the state energies in four digitized STA experiments.
The square, circle, triangle and diamond stand for the total number $M=1,2,3,4$ of digitized steps, respectively.}
\label{fig_02}
\end{figure}

In the first step, we explore the ground state $|\varphi_0(H)\rangle$ at $R=0.05$ \AA, where the energy parameters
are $g_0\!=\!10.08~E_\mathrm{h}$, $g\!=\!-1.055~E_\mathrm{h}$ and $g_{12}\!=\!0.1557~E_\mathrm{h}$ ($E_\mathrm{h}\!\approx\!27.2$ eV)~\cite{CollessPRX18}.
The initial Hamiltonian is selected to be $H_0 = g (Y_A + Y_B)$ and the adiabatic Hamiltonian $H_\mathrm{ad}(t)=H_0+\lambda(t)(H-H_0)$ is given by Eq.~(\ref{eq_00}).
The time-adjusting coefficient is set to be $\lambda(0\le t\le T)=\sin^2(\pi t/2T)$ and this form is applied to all the experiments
and numerical simulations in this paper.
The initial ground state $|\varphi_0(H_0)\rangle=(|0\rangle+i|1\rangle)_A(|0\rangle+i|1\rangle)_B/2$ is experimentally generated
by applying the $R_{X_A}(-\pi/2)$ and $R_{X_B}(-\pi/2)$ gates onto the two-qubit state $|00\rangle$.
Due to the condition of $|g_{12}|\!\ll\!|g|$, the two-qubit coupling is separated into $Y_AY_B\approx Y_A+Y_B$ under
the single-qubit approximation (see Appendix~\ref{appA})~\cite{chenxiPRAPP20}. Accordingly, the counter-diabatic Hamiltonian is approximated as
%\be
$H_\mathrm{cd}(t) \approx g_\mathrm{cd}(t)(X_A+X_B)$
%\label{eq_06}
%\ee
with $g_\mathrm{cd}(t) = \dot{\lambda}(t)/[2\zeta^2(t)+2\lambda^2(t)]$ and $\zeta(t)=1-(1-g_{12}/g)\lambda(t)$.
With both the adiabatic part $H_\mathrm{ad}(t)$ and the counter-diabatic part $H_\mathrm{cd}(t)$,
the total Hamiltonian is organized into
\be
H_\mathrm{tot}(t) &\approx& \lambda(t)g_0+\frac{\omega_1(t)}{2}Y_AY_B+\frac{\omega_2(t)}{2}(Y_A+Y_B) \no \\
&&+\frac{\omega_3(t)}{2}(X_A+X_B)+\frac{\omega_4(t)}{2}(Z_A+Z_B),
\label{eq_07}
\ee
with $\omega_1(t) = 2\lambda(t)g_{12}$, $\omega_2(t) = 2[1-\lambda(t)]g$, $\omega_3(t) = 2g_\mathrm{cd}(t)$ and $\omega_4(t) = 2\lambda(t)g$.
The STA operation time is empirically selected at $g T\approx 0.8$, which is much shorter than that for an adiabatic process ($gT_\mathrm{min}\approx9$).
%The operator sequence in Eq.~(\ref{eq_07}) is empirically selected for the best performance of the digitized STA.
The whole time evolution of the $M$-step digitized STA is given by Eq.~(\ref{eq_02})
and each $m$-th partial time evolution operator is set to be
\be
U_{m} &=& R_{Z_A}(\theta_{m; 4})R_{Z_B}(\theta_{m; 4})R_{X_A}(\theta_{m; 3})R_{X_B}(\theta_{m; 3}) \no \\
&& \times R_{Y_A}(\theta_{m; 2})R_{Y_B}(\theta_{m; 2})R_{Y_AY_B}(\theta_{m; 1}),
\label{eq_08}
\ee
with
\be
\left\{\ba{cclcl} \theta_{m, 1} &=& 2\lambda(m\Delta t)g_{12}\Delta t \\
\theta_{m, 2} &=& 2[1-\lambda(m\Delta t)]g \Delta t \\
\theta_{m, 3} &=& 2g_\mathrm{cd}(m\Delta t)\Delta t \\
\theta_{m, 4} &=& 2\lambda(m \Delta t)g\Delta t
\ea
\label{eq_09}
\right..
\ee
The two-qubit rotation $R_{Y_AY_B}(\theta)$ in Eq.~(\ref{eq_08}) is experimentally realized by~\cite{BarendsNC15}
\be
R_{Y_AY_B}(\theta)=R_{X_A}(\pi/2)R_{X_B}(\pi/2)U_\mathrm{CN}R_{Z_B}(\theta)U_\mathrm{CN}R_{X_A}(-\pi/2)R_{X_B}(-\pi/2),
\label{eq_10}
\ee
where $U_\mathrm{CN}$ stands for a two-qubit CNOT gate.
The corresponding pulse sequence is shown in Fig.~\ref{fig_02}(a). In practice, we perform the digitized STA experiment up to four steps ($1\!\le\!M\!\le\!4$).

%{The rotation angles $\{\theta_{m; j}\}$ are theoretically determined by the time-varying parameters $\{\omega_j(t)\}$,
%and experimentally adjusted to improve the accuracy of $U_m$.
%and measure the state evolution with the quantum state tomography (QST) during each step.}

\begin{figure}[tp]
\centering
 \includegraphics[width=0.5\columnwidth]{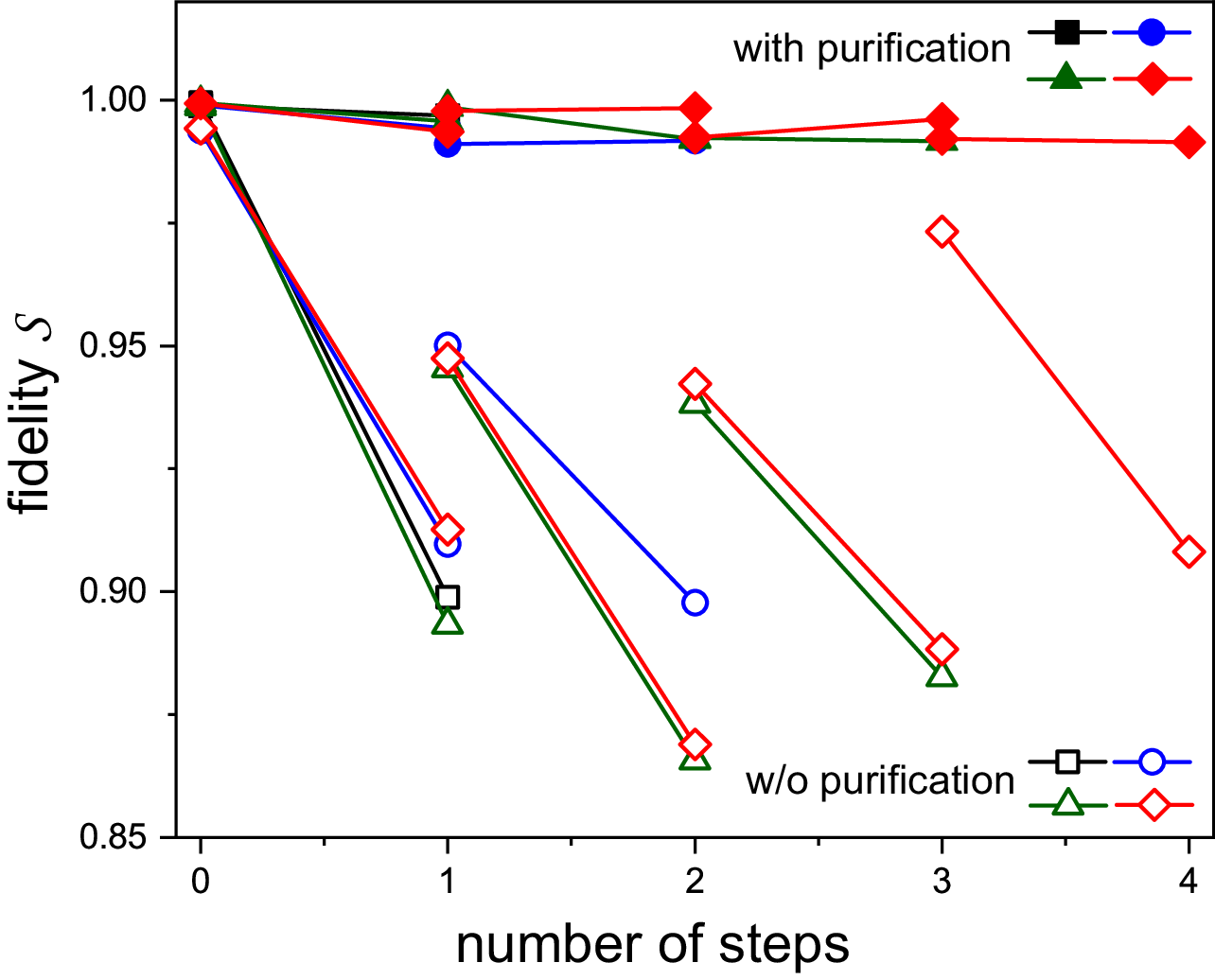}
\caption{The fidelity of the experimental density matrix $\rho_\mathrm{exp}(t)$
in comparison to the theoretical simulation $|\Psi_\mathrm{theo}(t)\rangle\langle \Psi_\mathrm{theo}(t)|$
for the experiment shown in Fig.~\ref{fig_02}. The open symbols represent the results without the state purification.
The solid symbols represent the results by selecting the pure state $|\psi_i(t)\rangle$ with the largest population
after the diagonalization of $\rho_\mathrm{exp}(t)$. The square, circle, triangle and diamond stand for
the total number $M=1,2,3,4$ of digitized steps, respectively.}
\label{fig_03}
\end{figure}

In our experiment on superconducting qubits, errors occur during the initialization, operation and measurement.
For example, the input pulse is generated by a digital-analog-converter (DAC) so that the pulse experienced by
qubits can deviate from the designed shape due to the filtering effect. A single STA step in Eq.~(\ref{eq_08}) involves 11 single-qubit gates and 2 two-qubit gates
and the operation error can quickly accumulate through an $M$-step STA process. Due to the nature of a transmon qubit,
the error is unavoidable in the state measurement. In addition to a careful calibration, we introduce three approaches to further
correct errors and achieve a reliable output. (1) At each $m$-th step, the starting state $|\Psi((m-1)\Delta t)\rangle$ is
re-prepared by a two-qubit universal quantum circuit (4 rotation gates and 1 CNOT gate)~\cite{ShendePRA03} after it is experimentally determined in the previous $(m-1)$-th step.
Our modification is equivalent to an evolution of $|\Psi(0)\rangle\rightarrow|\Psi((m-1)\Delta t)\rangle\rightarrow|\Psi(m\Delta t)\rangle$,
which can largely suppress the error accumulation in the pathway of $|\Psi(0)\rangle\rightarrow|\Psi(\Delta t)\rangle\rightarrow\cdots\rightarrow|\Psi((m-1)\Delta t)\rangle$.
For an $N(>2)$-qubit system, the universal quantum circuit can become difficult but a single-step production of $|\Psi(0)\rangle\rightarrow|\Psi((m-1)\Delta t)\rangle$
is practically possible. In the VQE protocol, the state determination of $|\Psi(0)\rangle\rightarrow|\Psi(H)\rangle$ is a single-step operation
although the rotation angles vary step by step~\cite{PeruzzoNC14,MalleyPRX16,KandalaNat17,GoogleSci20,CollessPRX18}.
(2) At each $m$-th step, the rotation angles $\{\theta_{m; j}\}$ in Eq.~(\ref{eq_09}) are adjusted to improve the reliability of $U_m$
since the overall effect of various errors behaves as a `phase error' in our generalized rotation operators.
The angle adjustment is in general a small value ($\lesssim 5^\circ$) around its theoretical design,
and a similar experimental technique can be found in a previous VQE experiment of the hydrogen molecule~\cite{MalleyPRX16}.
(3) Due to the limitation of our current experimental setup, the previous two approaches cannot fully remove the errors
so that we apply a state purification approach as follows. For an arbitrary mixed state, the density matrix can be diagonalized
and expanded into $\rho_\mathrm{exp}(t)=\sum_i P_i(t) |\psi_i(t)\rangle\langle \psi_i(t)|$, where $|\psi_i(t)\rangle$ is
a pure state and $P_i(t)$ is its population.
We expect that the state evolution is not far away from the theoretical simulation.
The state $|\psi_i(t)\rangle$ with the largest population is thus considered to be an error-corrected experimental result.
To visualize the influence of the error correction, we sequentially measure the starting and ending density matrices,
$\rho((m-1)\Delta t)$ and $\rho(m\Delta t)$, with the quantum state tomography (QST).
Next we calculate a fidelity function,
\be
\mathcal S = \langle \Psi_\mathrm{theo}(t)|\rho_\mathrm{exp}(t)|\Psi_\mathrm{theo}(t)\rangle,
\label{eq_11}
\ee
between the experimental result $\rho_\mathrm{exp}(t)$ and the theoretical simulation
$|\Psi_\mathrm{theo}(t)\rangle\langle \Psi_\mathrm{theo}(t)|$ at the starting time $t=(m-1)\Delta t$
and the ending time $t=m\Delta t$. The propagation results of $\mathcal S$ for the $M$-step digitized
STA ($1\le M\le 4$) are shown in Fig.~\ref{fig_03}.
Except for $|\Psi(0)\rangle$, the starting state $|\Psi((m-1)\Delta t)\rangle$ generated
by the universal quantum circuit  exhibits a $\sim 5\%$ error,
and a one-step digitized STA results in an extra $5\%\sim10\%$ error
for the ending state $|\Psi(m\Delta t)\rangle$.
After the purification, the state fidelities of $\mathcal S$ are consistently
improved to be $>99\%$.  The above three treatments are also used in other experiments in this paper,
but may be unnecessary with the future improvement of sample quality and control accuracy.

Next we present the time evolution of the error-corrected experimental result $|\Psi(t\!=\!m\Delta t)\rangle$
in comparison to the exact ground state $|\varphi_{0}(H)\rangle$. Similarly, another fidelity function,
\be
\mathcal F_m\!=\!|\langle\varphi_{0}(H)|\Psi(t\!=\!m\Delta t)\rangle|^2,
\ee
is introduced for a quantitative description, and the experimental results of $\mathcal F_m$ are presented in Fig.~\ref{fig_02}(b).
For $M\ge2$, the digitized STA efficiently drags the initial state ($\mathcal F_0=23\%$) to
the exact state (the best fidelity $\mathcal F_M\!\approx\!99\%$).
%Although the Trotter-Suzuki decomposition implicitly requires a large number of $M$, this fast convergence could be attributed to a weak two-qubit interaction.
%($|g_{12}|\!\ll\!|g|$ at $R\!=\!0.05$ \AA).
We also calculate the state energies
$E_m\!=\!\langle \Psi(m\Delta t)|H|\Psi(m\Delta t)\rangle$
through each $M$-step process [see Fig.~\ref{fig_02}(c)],
the experimental extraction is almost the same as the exact value $\varepsilon_0(R\!=\!0.05~\mathrm{\AA})\!=\!7.96$ $E_\mathrm{h}$.
%{\color{red}the experimentally determined value of the ground state energy is almost the same.}

\begin{figure}[tp]
\centering
 \includegraphics[width=0.7\columnwidth]{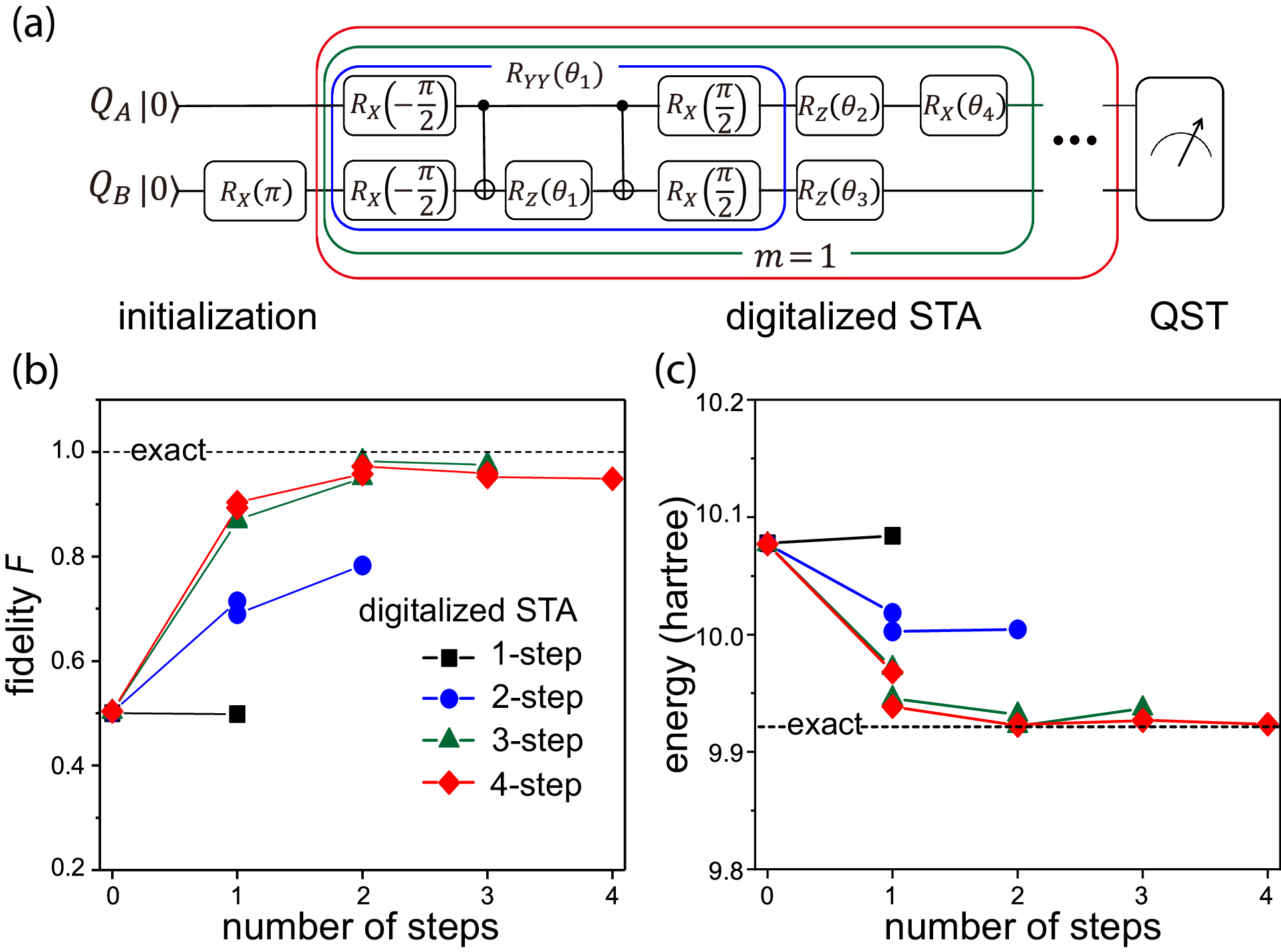}
\caption{Determination of the first excited state of H$_2$ via a digitized STA eigensolver for the internuclear distance at $R=0.05$ \AA.
(a) A quantum circuit diagram of the $M$-step digitized STA process.
(b) The evolution of the state fidelities in four digitized STA experiments ($1\le M\le 4$)
in comparison with the exact result. (c) The evolution of the state energies in four digitized STA experiments.
The square, circle, triangle and diamond stand for the total number $M=1,2,3,4$ of digitized steps, respectively.}
\label{fig_04}
\end{figure}

In the second step, we explore the first excited state $|\varphi_{n=1}(H)\rangle$ at $R\!=\!0.05~\mathrm{\AA}$ while the other two excited states %in Eq.~(\ref{eq_03})
can be investigated similarly. The initial Hamiltonian is changed to be $H_0\!=\!gZ_A$,
and the initial state $|\Psi(0)\rangle\!=\!|01\rangle$
is obtained by applying the $R_{X_B}(\pi)$-gate onto the state $|00\rangle$.
Following the same procedure as above, a short operation time is selected at $g T\approx 4.0$
and each $m$-th partial time evolution operator in the digitized STA is given by
\be
U_{m} = R_{X_A}(\theta_{m; 4})R_{Z_A}(\theta_{m; 3})R_{Z_B}(\theta_{m; 2})R_{Y_AY_B}(\theta_{m; 1}).
\label{eq_12}
\ee
with
\be
\left\{\ba{cclcl} \theta_{m, 1} &=& 2\lambda(m\Delta t)g_{12}\Delta t \\
\theta_{m, 2} &=& 2g \Delta t \\
\theta_{m, 3} &=&2\lambda(m \Delta t)g\Delta t \\
\theta_{m, 4} &=&2g_\mathrm{cd}(m\Delta t)\Delta t
\ea
\label{eq_13}
\right.
\ee
and $g_\mathrm{cd}(t) = -\dot{\lambda}(t)gg_{12}/[2g+2g_{12}\lambda^2(t)]$. %with $c=g/g_{12}$ %and $g_\mathrm{cd}(t)= -g g_{12}\dot{\lambda}(t)/[2\lambda^2(t)g^2_{12}+2g^2]$.}}
The corresponding pulse sequence is plotted in Fig.~\ref{fig_04}(a). The experimental results
of the state fidelity $\mathcal F_m\!=\!|\langle\varphi_{n=1}(H)|\Psi(m\Delta t)\rangle|^2$
and the state energy $E_m$ are presented in Figs.~\ref{fig_04}(b) and \ref{fig_04}(c).
Consistent with the behavior in the ground state, the
digitized STA efficiently drags the initial state ($\mathcal F_0\!=\!50\%$) to
the exact first excited state (the best fidelity $\mathcal F_M\!\approx\!98\%$) for $M\!\ge\!3$,
leading to an excellent prediction of the eigenenergy at $\varepsilon_1(R\!=\!0.05~\mathrm{\AA})\!=\!9.92$ $E_\mathrm{h}$.

\begin{figure}[tp]
\centering
 \includegraphics[width=0.7\columnwidth]{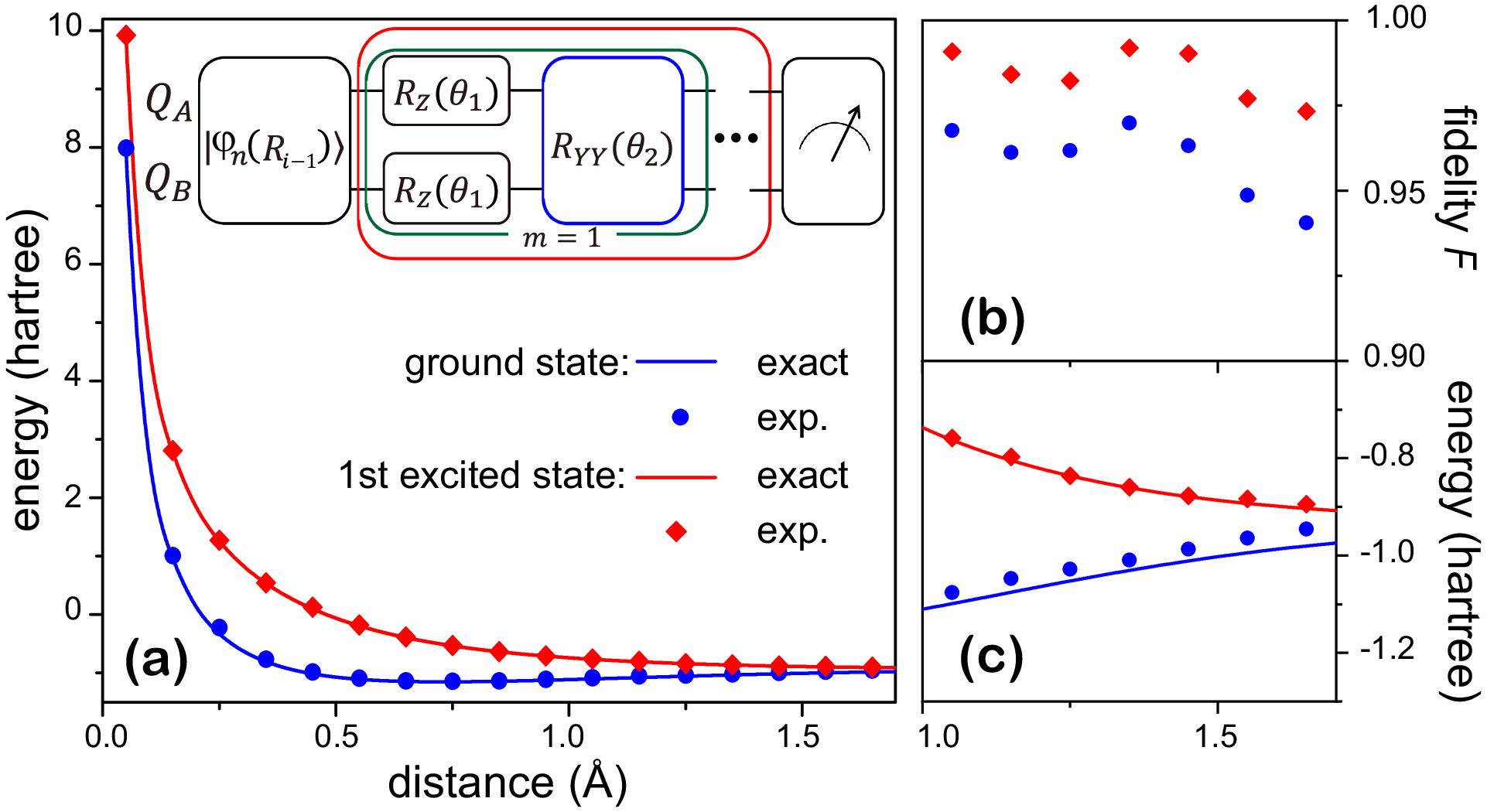}
\caption{Experimental determination of the potential energy landscapes for both the ground and excited states of H$_2$ via the digitized STA and sequential digitized adiabatic eigensolvers.
(a) The two eigenergies ($\varepsilon_{n=0, 1}$) as the functions of the internuclear distance $R$.
The quantum circuit diagram of a sequential digitized adiabaticity is shown in the inset. In the large-distance range ($1.05~\mathrm{\AA}\!\le\!R\!\le\!1.65~\mathrm{\AA}$),
the experimental results of the state fidelities and eigenenergies are enlarged in (b) and (c).
The symbols (circles and diamonds) represent the experimental results while the solid lines represent the exact results from the theoretical calculation.}
\label{fig_05}
\end{figure}

In the final step, we explore the electronic states of H$_2$
for various internuclear distances $R$, i.e., $|\varphi_{n=0, 1}(R)\rangle\!\equiv\!|\varphi_{n=0, 1}(H(R))\rangle$.
With the increase of $R$, the amplitude of $g(R)$ gradually decreases and  eventually becomes smaller
than $g_{12}(R)$. For the two initial Hamiltonians, $H_0=g(Y_A+Y_B)$ and $H_0=gZ_A$,
the previous single-qubit approximation gradually fails. Instead of an expansion over nested commutators~\cite{PWPRL19},
we apply a sequential digitized adiabatic approach. In the  range of $0.05~\mathrm{\AA}\!\le\!R\!\le\!1.65~\mathrm{\AA}$,
a sequence of the internuclear distances is set as $\{R_0=0.05~\mathrm{\AA}, R_1=0.15~\mathrm{\AA}, \cdots, R_{16}=1.65~\mathrm{\AA}\}$,
which satisfies $R_{i}\!=\!R_{i-1}+\Delta R$ and $\Delta R=0.1~\mathrm{\AA}$. To obtain the electronic structure at the $i$-th distance,
the initial Hamiltonian is selected to be $H_0=c(R_{i})H(R_{i-1})$ with $c(R_{i})=g_{12}(R_{i})/g_{12}(R_{i-1})$ and the initial states
are the corresponding ground and first excited states, $|\Psi(0)\rangle\!=\!|\varphi_{n=0,1}(R_{i-1})\rangle$,   %=|\varphi_n(H(R_{i-1})\rangle$,
which have been determined at the previous $(i-1)$-th distance. The adiabatic Hamiltonian is given accordingly by
$H_\mathrm{ad}(t)=c(R_i)H(R_{i-1})+\lambda(t)[H(R_i)-c(R_i)H(R_{i-1})]$. Due to a relatively small deviation between
$|\varphi_{n}(R_{i-1})\rangle$ and $|\varphi_n(R_i)\rangle$, the adiabatic evolution can converge quickly
and we practically choose a digitized adiabatic process to realize the eigensolver.
In detail, the time evolution operator of the digitized adiabaticity is set to be $U_{\mathrm{ad}}\!=\!U_M\cdots U_m\cdots U_1$ and
\be
U_{m=1,\cdots,M} = R_{Y_AY_B}(\theta_{m;2})R_{Z_A}(\theta_{m;1})R_{Z_B}(\theta_{m;1}).
\label{eq_14}
\ee
The rotation angles are $\theta_{m;1}=2g_\lambda(m\Delta t) \Delta t$ and $\theta_{m;2}=2g_{12}(R_i)\Delta t$
where we introduce a function, $g_\lambda(t)\!=\!\lambda(t)g(R_{i})\!+\![1\!-\!\lambda(t)]c(R_{i})g(R_{i-1})$.
The pulse sequence of this digitized adiabatic evolution is shown in the inset of Fig.~\ref{fig_05}(a).
The operation time is in the range of $1.0\le gT\le 1.6$.
For the ground state $|\varphi_0(R_i)\rangle$, the number of digitized steps are $M=2$ for $0.15~\mathrm{\AA}\!\le\!R_i\!\le\!1.05~\mathrm{\AA}$
and $M=5$ for $1.15~\mathrm{\AA}\!\le\!R_i\!\le\!1.65~\mathrm{\AA}$.
For the first excited state $|\varphi_1(R_i)\rangle$, the number of digitized steps is always $M=2$.
To reduce the accumulation of experimental errors, the initial states
$|\varphi_{n}(R_{i-1})\rangle$ are re-prepared according to the numerical simulation result (including a systematic error from
the sequential digitized adiabaticity) with the universal quantum circuit~\cite{ShendePRA03}.
The experimental result of the sequential digitized adiabaticity is summarized in Fig.~\ref{fig_05}(a)-\ref{fig_05}(c).
In the most relevant range around the equilibrium distance, the state fidelities
of both the ground and first excited states satisfy $\calF_M\gtrsim 98\%$. In the range of large distances
($1.05~\mathrm{\AA}\!\le\!R_i\!\le\!1.65~\mathrm{\AA}$), the fidelity of the ground state is $\calF_M\gtrsim 94\%$
while that of the first excited state is $\calF_M\gtrsim 97\%$ [see Fig.~\ref{fig_05}(b)]. The fidelity drop
is expected to slow down with the decrease of the distance deviation $\Delta R$.
The potential energy landscape extracted experimentally is plotted in Fig.~\ref{fig_05}(a),
showing an excellent agreement with the theoretical prediction.
An enlarged view of the energy landscape for $1.05~\mathrm{\AA}\!\le\!R_i\!\le\!1.65~\mathrm{\AA}$ is shown in Fig.~\ref{fig_05}(c).
We thus obtain an accurate description of the electronic structure of H$_2$ using a delicate and efficient design
of the digitized STA and the sequential digitized adiabaticity.

%(In addition to the extension of the single-qubit approximation, the rearrangement of the initial Hamiltonian shows its
%advantage in the convergence speed.)

\subsection{Topological BHZ model}
\label{sec_032}

The BHZ model can be used to describe the quantum spin Hall effect in a two-dimensional square lattice (the lattice constant is $a$)~\cite{ZhangSci06}.
For simplicity, we consider one $s$-orbital  and one $p$-orbital at each lattice point. In addition to the intra-orbital interactions, the spin-modulated
and orientation-modulated inter-orbital interactions are included for the nearest neighbor hopping. After a discrete Fourier transform,
the tight-binding Hamiltonian in the momentum space is block diagonal against the wavevector $\vec{k}$, giving $H=\sum_{\vec{k}}H(\vec{k})$ and
\be
H(\vec{k})=\sum_{\gamma_{1}=s,p}\sum_{\gamma_2=s,p}\sum_{\sigma =\pm}|\vec{k}\gamma_1\sigma\rangle [H(\vec{k})]_{\gamma_1\sigma ;\gamma_2\sigma } \langle \vec{k}\gamma_2\sigma |.
\label{eq_16}
\ee
By mapping the two orbitals $\{|s\rangle, |p\rangle\}$ onto a two-state spin, Eq.~(\ref{eq_16}) is reorganized into   %a two-qubit form,
\be
H(\vec{k}) =  g_0(\vec{k})+g(\vec{k})Z_{B}+\left[g_{12}(k_x)X_{B}+g_{12}(k_y)Y_{B}\right]Z_{A},
\label{eq_17}
\ee
where the subscripts $A$ and $B$ denote the spin and orbital degrees of freedoms, respectively.
The coefficients are explicitly given by
\be
\left\{\ba{ccl} g_0(\vec{k}) &=&C_1 f(\vec{k})  \\
g(\vec{k}) &=& C_2 f(\vec{k})-C_3  \\
g_{12}(h=k_x, k_y) &=& C_4 \sin(h a)
\ea
\label{eq_18}
\right.
\ee
with $f(\vec{k}) = 8[\sin^2(k_x a/2)\!+\!\sin^2(k_y a/2)]$.
In our experiment, the three parameters are set to be $C_1=0.020$ eV, $C_2=0.027$ eV, $C_3=0.055$ eV and $C_4=0.073$ eV
from the HgTe/CdTe semiconductor quantum well~\cite{JiangPRB09}.
The symmetry of $H$ indicates two-fold degenerate energy bands and a Dirac point at the $\Gamma$-point ($k_x\!=\!k_y\!=\!0$) of the first
Brillouin zone (FBZ). To demonstrate the reliability of the digitized STA and the sequential digitized adiabaticity, we drive the two-qubit
system to determine the valence and conduction bands along the X-$\Gamma$-X linecut ($-\pi\!\le\!k_x a\!<\!\pi$ and $k_y\!=\!0$) in the FBZ.

\begin{figure}[tp]
\centering
 \includegraphics[width=0.75\columnwidth]{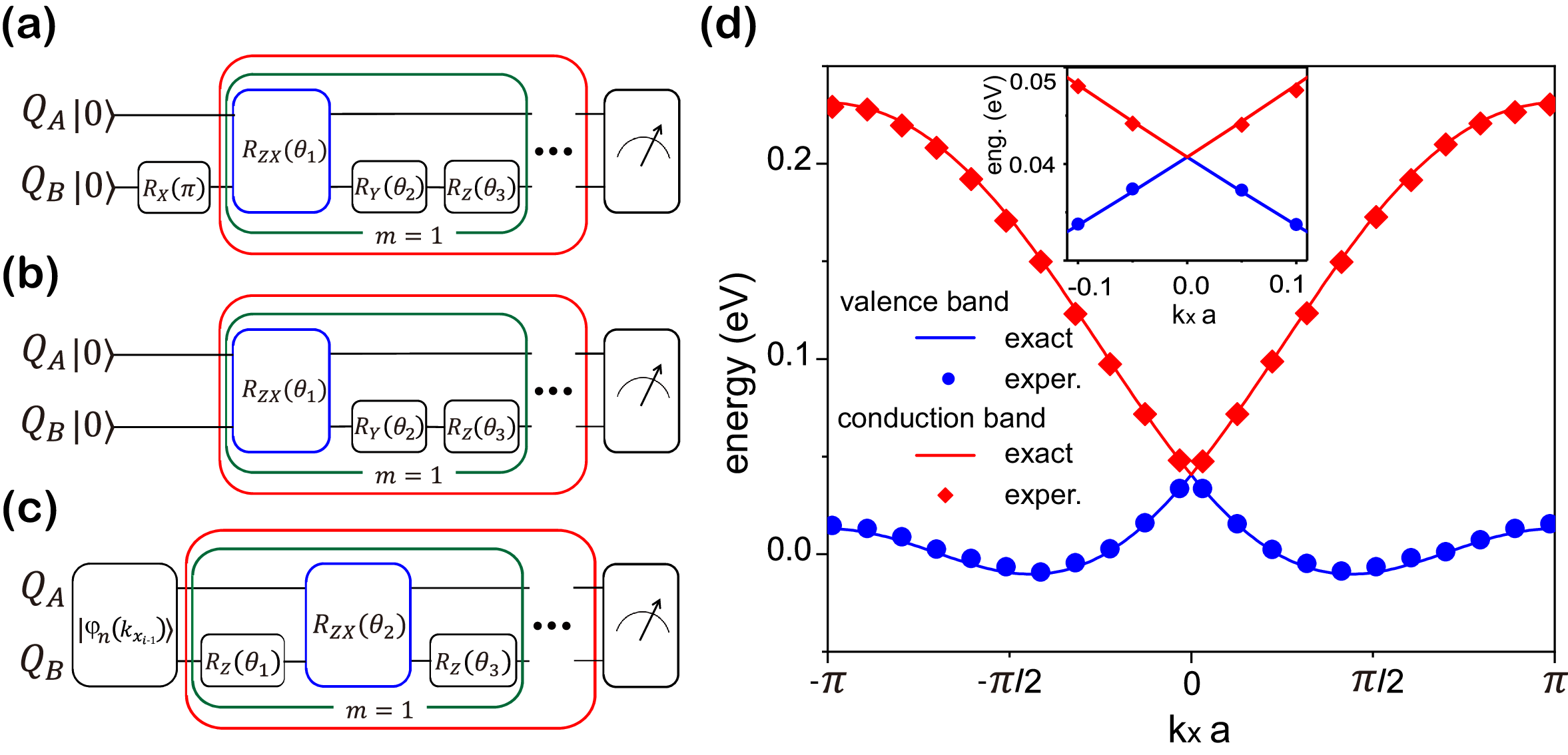}
\caption{Experiment of the BHZ model. (a) and (b) The quantum circuit diagrams
of the digitized STA to determine the ground and excited eigenstate at the starting wavevectors, $k_{x_{0}}=\pm0.1a^{-1},\pm3.1a^{-1}$.
(c) The quantum circuit diagram of the sequential digitized adiabaticity for other wavevectors, $k_{x}\neq k_{x_{0}}$.
(d) The experimental determination of the valence and conduction bands along the X-$\Gamma$-X linecut.
Inset is an enlarged view of (d) around the Dirac point.
The symbols (circles and diamonds) represent the experimental results while the solid lines represent the exact results from the theoretical calculation. }
\label{fig_06}
\end{figure}

Similar to the treatment in H$_2$, we first explore the electronic states $|\varphi_{n=0, 1}(k_{x_0})\rangle\!\equiv\!|\varphi_n(H(k_x\!=\!k_{x_0},k_y\!=\!0))\rangle$
at two sets of starting points: one near the $\Gamma$-point ($k_{x_0}\!=\!\pm0.1a^{-1}$)
and the other near the X-point ($k_{x_0}\!=\!\pm3.1a^{-1}$). For each wavevector, the initial Hamiltonian is selected to be $H_0 = g(k_{x_0})Z_B$.
The initial ground state of $H_0$ is $|\varphi_0(H_0)\rangle\!=\!|01\rangle$, generated by the $R_{X_B}(\pi)$ gate onto the two-qubit state $|00\rangle$,
while the initial excited state is $|\varphi_1(H_0)\rangle\!=\!|00\rangle$.
With respect to the adiabatic Hamiltonian, $H_\mathrm{ad}(t)=H_0+\lambda(t)(H-H_0)$,
the counter-diabatic Hamiltonian $H_\mathrm{cd}(t)$ is obtained under the single-qubit approximation, which is valid with the consideration of $|g_{12}(k_{x_0})|\ll |g(k_{x_0})|$.
In the $M$-step digitized STA, the partial time evolution operator at each step is given by
\be
U_{m} &=& R_{Z_B}(\theta_{m; 3})R_{Y_B}(\theta_{m; 2})R_{Z_AX_B}(\theta_{m; 1}).
\label{eq_19}
\ee
with
\be
\left\{\ba{ccl}
\theta_{m; 1}&=&2\lambda(m\Delta t)g_{12}(k_{x_0})\Delta t \\
\theta_{m; 2}&=&2g_\mathrm{cd}(k_{x_0};m\Delta t)\Delta t\\
\theta_{m; 3}&=&2g(k_{x_0})\Delta t
\ea
\label{eq_20}
\right.
\ee
and $g_\mathrm{cd}(k_{x_0};t) = \dot{\lambda}(t)g(k_{x_0})g_{12}(k_{x_0})/[2g(k_{x_0})+2g_{12}(k_{x_0})\lambda^2(t)]$.
The corresponding pulse sequences for both the ground and excited states are
shown in Figs.~\ref{fig_06}(a) and \ref{fig_06}(b), respectively.
In practice, we perform eight 4-step digitized STA experiments ($0.5\le gT\le 0.6$) and determine the
corresponding eigenstates.
For both $k_{x_0}\!=\!\pm0.1a^{-1}$ and $\pm3.1a^{-1}$, the experimental state fidelities of $|\varphi_{n=0, 1}(k_{x_0})\rangle$
are around $98\%\sim 99\%$.

Next we explore the electronic states  $|\varphi_{n=0, 1}(k_x)\rangle$ at different wavevectors along the X-$\Gamma$-X linecut
through the sequential digitized adiabaticity. With respect to each starting wavevector ($k_{x_0}=\pm 0.1 a^{-1}, \pm 3.1 a^{-1}$),
a sequence of wavevectors, $\{k_{x_0}, k_{x_1}, \cdots, k_{x_i}, \cdots\}$, is assigned.
For each sequence, the ending wavector is at $k_x=\pm 1.6 a^{-1}$ and the wavevector deviation is $\Delta k_x=\pm 0.3 a^{-1}$.
In addition, we investigate two small wavevectors, $k_x=\pm 0.05 a^{-1}$, using the reference points at $k_{x_0}=\pm 0.1 a^{-1}$.
At each $i$-th wavevector $k_{x_i}$,
the initial Hamiltonian is set to be $H_0=c(k_{x_i})H(k_{x_{i-1}})$ with $c(k_{x_i})=g_{12}(k_{x_i})/g_{12}(k_{x_{i-1}})$,
while the initial state is $|\Psi(0)\rangle\!=\!|\varphi_{n=0,1}(k_{x_{i-1}})\rangle$. The adiabatic Hamiltonian
is given by $H_\mathrm{ad}(t)=c(k_{x_i})H(k_{x_{i-1}})+\lambda(t)[H(k_{x_i})-c(k_{x_i})H(k_{x_{i-1}})]$.
In the $M$-step digitized adiabatic process, each partial time evolution operator is given by
\be
U_m = R_{Z_B}(\theta_{m;3})R_{Z_AX_B}(\theta_{m;2})R_{Z_B}(\theta_{m;1})
\label{eq_21}
\ee
with $\theta_{m;1}= 2\Delta g(k_{x_i})\lambda(m\Delta t)\Delta t$, $\theta_{m;2}=2g_{12}(k_{x_i})\Delta t$ and $\theta_{m;3} = 2c(k_{x_i})g(k_{x_{i-1}})\Delta t$.
Here we introduce a deviation function, $\Delta g(k_{x_i})\!=\!g(k_{x_i})\!-\!c(k_{x_i})g(k_{x_{i-1}})$.
The pulse sequence of the digitized adiabaticity is shown in the inset of Fig.~\ref{fig_06}(c).
The separation of $R_{Z_B}(\theta_{1})$ and $R_{Z_B}(\theta_{3})$ in Eq.~(\ref{eq_21})
is found to yield a better performance than the combined case.
The operation times are selected in the range of $0.4\le gT\le 1.8$.
For $|k_{x}|\!\le\!1.0a^{-1}$, the number of adiabatic steps is $M=2$,
while for $1.3 a^{-1}\!\le\!|k_{x}|\!\le\!2.8 a^{-1}$, this number is changed to be $M=5$.
The experimental state fidelities of both the valence and conduction bands along the X-$\Gamma$-X linecut
satisfy $\calF_M\gtrsim 98\%$. In Fig.~\ref{fig_06}(d), we plot the corresponding
experimental band structures, $\varepsilon_{n=0, 1}(k_x)\!=\!\langle\varphi_{n}(k_x)|H|\varphi_n(k_x)\rangle$,
which agrees excellently with the theoretical prediction.
In the inset of Fig.~\ref{fig_06}(d), we experimentally extract a linear dispersion relation,
$\varepsilon_{n}(k_x)-2C_1\!\sim\!\pm C_4k_xa$ around the Dirac point ($k_x\!=\!0$).

\section{Numerical Simulation of Hydrogen Chains}
\label{sec3c}

Our experiments of the two-atom hydrogen molecule and the BHZ model have revealed the applicability of the digitized STA
followed by the sequential digitized adiabaticity in the determination of electronic states.
Next we investigate the ground state for $N(>2)$-atom hydrogen chains~\cite{GoogleSci20}. Due to the restriction of our current setup,
it is difficult for us to reliably perform an $N(>2)$-qubit experiment. Therefore, we only present the results of numerical simulation
%via the digitized STA and sequential one-step adiabaticity
in this paper and the experimental implementation will be performed in the future.

\begin{figure}[tp]
\centering
 \includegraphics[width=0.65\columnwidth]{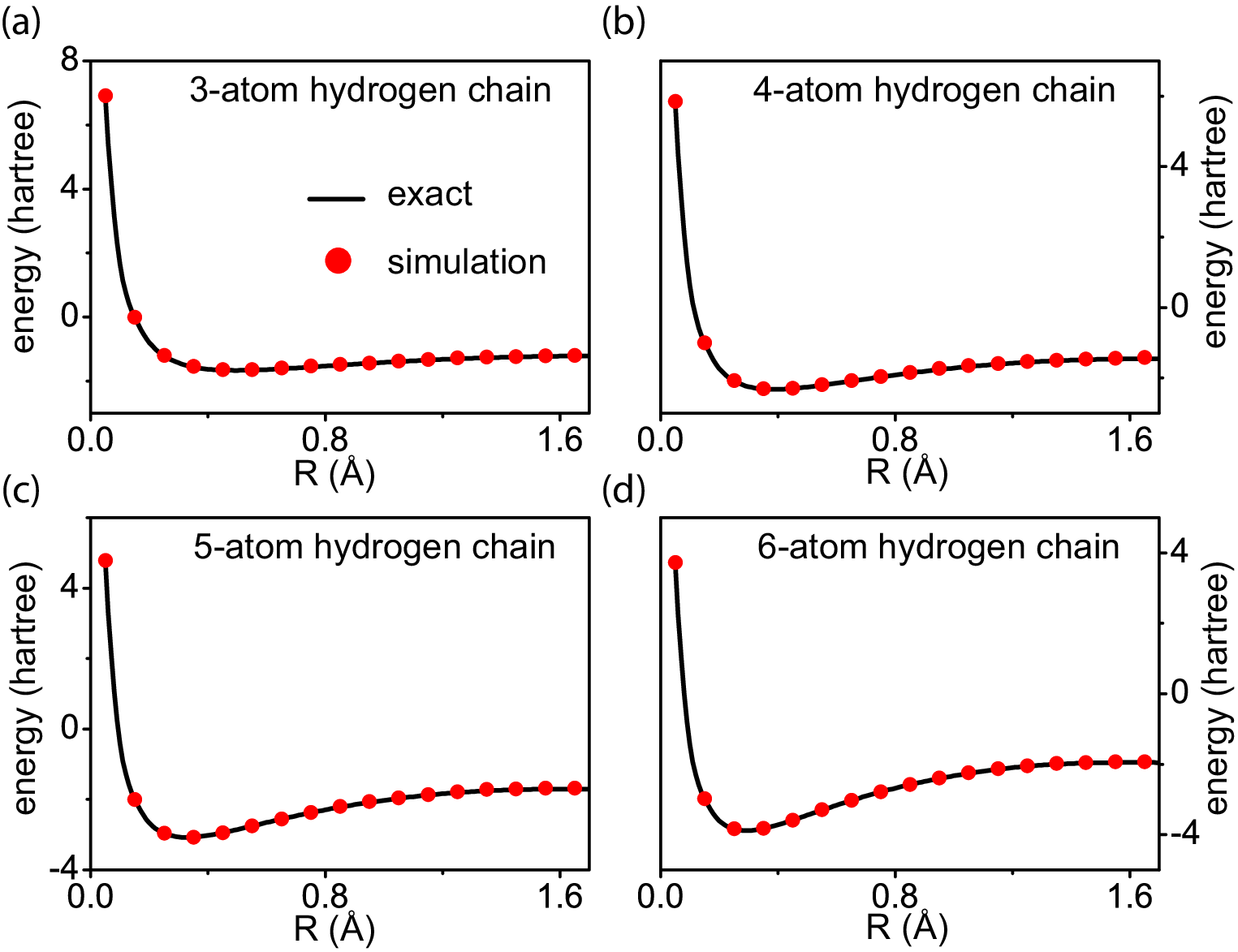}
\caption{The ground state energy landscapes of the 3-, 4-, 5- and 6-atom hydrogen chains,
where $R$ is the internuclear distance between two adjacent atoms.
The red circles are obtained from the numerical simulation of the digitized STA and the sequential digitized adiabaticity while the black solid lines are the exact results.}
\label{fig_07}
\end{figure}

The Hamiltonian of an $N$-atom hydrogen chain is written as
\be
H(N; R) = g_0(R)+\sum_{i=1}^N g(R) Z_i +\sum_{i=1}^{N-1} g_{12}(R) Y_i Y_{i+1},
\ee
where $R$ is the internuclear distance between each pair of two adjacent hydrogen atoms.
For simplicity, the $R$-dependencies of $g_0(R)$, $g(R)$ and $g_{12}(R)$ are assumed to be the same as those in the two-atom hydrogen molecule.
(1) For the specified distance ($R\!=\!R_0\!=\!0.05$ \AA), the initial Hamiltonian is set to be
$H_0=H(N-1; R_0)+g(R_0) Z_N$ and the adiabatic Hamiltonian reads $H_\mathrm{ad}(t)\!=\!H(N-1; R_0)+g(R_0) Z_N+\lambda(t)g_{12}(R_0)Y_{N-1}Y_N$.
The initial state is set to be $|\Psi(0)\rangle=|\varphi_0(N-1; R_0)\rangle\otimes|0\rangle$,
where the ground state $|\varphi_0(N-1; R_0)\rangle$ of the $(N-1)$-atom chain is predetermined.
The adiabatic process is digitized into
\be
U_m = R_{Y_1Y_2}(\theta_{m; 3})\cdots R_{Y_{N-2}Y_{N-1}}(\theta_{m; 3})R_{Y_{N-1}Y_N}(\theta_{m; 2})R_{Z_1}(\theta_{m; 1})\cdots R_{Z_N}(\theta_{m; 1})
\ee
with $\theta_{m; 1}=2g(R_0)\Delta t$, $\theta_{m; 2}=2g_{12}(R_0)\lambda(m\Delta t)\Delta t$ and $\theta_{m; 3}=2g_{12}(R_0)\Delta t$.
(2) For the other distances ($R>R_0$), a sequential digitized adiabatic protocol is applied onto the same distance sequence,
$\{R_0=0.05~\mathrm{\AA}, R_1=0.15~\mathrm{\AA}, \cdots, R_{16}=1.65~\mathrm{\AA}\}$, as in Sec.~\ref{sec_031}.
At each $i$-th distance, the initial Hamiltonian is selected to be $H_0=c(R_{i})H(N;R_{i-1})$
and the adiabatic Hamiltonian is given accordingly by
$H_\mathrm{ad}(t)=c(R_i)H(N;R_{i-1})+\lambda(t)[H(N;R_i)-c(R_i)H(N;R_{i-1})]$.
The initial state $|\varphi_0(N; R_{i-1})\rangle$ is predetermined at the previous $(i-1)$-th distance.
For the digitized adiabatic operator, $U_{\mathrm{ad}}\!=\!U_M\cdots U_m\cdots U_1$, each partial time evolution is given by
\be
U_m= R_{Y_1Y_2}(\theta_{m; 2})\cdots R_{Y_{N-1}Y_N}(\theta_{m; 2})R_{Z_1}(\theta_{m; 1})\cdots R_{Z_N}(\theta_{m; 1})
\ee
with $\theta_{m;1}=2g_\lambda(m\Delta t) \Delta t$ and $\theta_{m;2}=2g_{12}(R_i)\Delta t$.
The adiabatic operation time is in the range of $0.1\le gT\le 0.9$ while the number of the digitized
steps is in the range of $3\le M\le 10$. In general, the value of $M$ increases with the increase of the chain length $N$
and the internuclear distance $R$.
The ground state energy landscapes ($\varepsilon_0$ versus $R$) obtained for the $N$-atom hydrogen chains ($3\!\le\!N\!\le\!6$)
are plotted in Figs.~\ref{fig_07}(a)-\ref{fig_07}(d). Through a series of sequential numerical simulations from the 3- to  6-atom chains,
we successfully determine the structure of the electronic ground states. The state fidelities from the numerical simulation
are $\gtrsim 99\%$ at short internuclear distances and still $\gtrsim 95\%$ for $1.05~\mathrm{\AA}\!\le\!R_i\!\le\!1.65~\mathrm{\AA}$.

\section{Summary}
\label{sec_04}

In this paper, we apply an eigensolver of the digitized STA and the sequential digitized adiabaticity to determine
the electronic states of various systems.
For the two-atom hydrogen molecule and the topological
BHZ model, the state determination is experimentally implemented in a two-qubit superconducting device.
(1) In the case of the hydrogen molecule, we select an internuclear distance at $R=R_0=0.05~\mathrm{\AA}$
with a weak inter-qubit interaction. Two uncoupled Hamiltonians are set to
be the initial Hamiltonians $H_0$ %which lead to the adiabatic Hamiltonian $H_\mathrm{ad}(t)=H_0+\lambda(t)(H-H_0)$
and the counter-diabatic Hamiltonians $H_\mathrm{cd}(t)$ are estimated under the single-qubit assumption.
Both the ground ($|\varphi_0(R=R_0)\rangle$) and first excited ($|\varphi_1(R=R_0)\rangle$) states are
experimentally determined by the 4-step digitized STA (state fidelities $\mathcal F\approx 99\%$ and $98\%$).
Subsequently, we set a sequence of internuclear distances,
$\{R_0=0.05~\mathrm{\AA}, R_1=0.15~\mathrm{\AA}, \cdots, R_{16}=1.65~\mathrm{\AA}\}$.
For each $i$-th distance $R_i$, the initial Hamiltonian is set to be $H(R_{i-1})$
at the $(i-1)$-th distance. With the initial states at $|\varphi_{0}(R_{i-1})\rangle$
and $|\varphi_1(R_{i-1})\rangle$, a $M$-step digitized adiabatic process
%of  $H_\mathrm{ad}(t)=c(R_i)H(R_{i-1})+\lambda(t)[H(R_i)-c(R_i)H(R_{i-1})]$
is  implemented to experimentally determine the ground and first excited states,
$|\varphi_{0}(R_{i})\rangle$ and $|\varphi_1(R_{i})\rangle$.
Through such sequential digitized adiabaticity, the potential energy landscape
of H$_2$ is extracted excellently with the state fidelities in the range of $94\%\sim99\%$.
(2) The same strategy is applied to the BHZ model to experimentally determine the valence and conduction bands along the X-$\Gamma$-X linecut ($k_y=0$) of the FBZ.
For the two sets of starting wavevectors $k_{x_0}$, the ground and excited states,
$|\varphi_{0}(k_{x_0})\rangle$ and $|\varphi_1(k_{x_0})\rangle$, are extracted by the 4-step digitized STA under the single-qubit approximation.
Through various sequences of wavevectors, $\{k_{x_0},\cdots, k_{x_i}, \cdots\}$, the electronic structure $|\varphi_{n=0,1}(k_{x_i})\rangle$
along the X-$\Gamma$-X linecut is obtained using the sequential digitized adiababticity, showing high state fidelities
($98\%\sim 99\%$). A linear dispersion relation,
$\delta\varepsilon_{n}(k_x)\!\propto\!k_x$, is also experimentally confirmed around the Dirac point ($k_x\!=\!0$).
(3) This eigensolver approach %of the digitized STA and the sequential one-step adiabaticity
is then extended to the $N$-atom hydrogen chains by a numerical simulation. With the predetermined
ground state $|\varphi_0(N=2; R_0)\rangle$ of the two-atom hydrogen molecule,
a series of sequential digitized adiabatic processes are applied. For the specified internuclear distance
$R=R_0$, the simulation procedure follows $|\varphi_{0}(N=2; R_0)\rangle\rightarrow|\varphi_{0}(N=3; R_0)\rangle\rightarrow\cdots$.
For other distances ($R>R_0$) at a given chain length ($N$), the simulation procedure follows
$|\varphi_{0}(N; R_0)\rangle\rightarrow|\varphi_{0}(N; R_1)\rangle\rightarrow\cdots$.
The overall state fidelities for the conditions of $3\le N\le 6$
and $0.05~\mathrm{\AA} \le R\le 1.65~\mathrm{\AA}$ are around $95\%\sim99\%$.

Our experiments and numerical simulation confirm the applicability of the digitized STA followed by the
sequential digitized adiabaticity. Conceptually speaking, this approach aims to determine the eigenstates $|\varphi_n\rangle$
of a quantum system in a parameter ($\Lambda-$) space. In the first step, we select one or a few starting points ($\Lambda_0$)
with weak inter-qubit interactions.
With nearly zero information about $|\varphi_n(\Lambda_0)\rangle$, a good strategy is to drag the system
from a simple eigenstate (e.g., a product state with respect to an uncoupled Hamiltonian) to the target eigenstate
via the digitized STA. Subsequently, we set sequences of $\{\Lambda_0, \Lambda_1, \cdots,\Lambda_i,\cdots\}$, to traverse
the parameter space. Due to a short state distance, the digitized adiabaticity can be applied to sequentially drag the system from
$|\varphi_n(\Lambda_{i-1})\rangle$ to $|\varphi_n(\Lambda_{i})\rangle$ with a high fidelity.
%In principle, we can
%extract the general $\Lambda$-dependence of the eigenstates $|\varphi_n(\Lambda)\rangle$ and the eigenenergies $\varepsilon_n(\Lambda)$.
This approach is in parallel with
the standard solver of a differential equation. For example, a time-dependent function $f(t)$ can be sequentially determined
by $f(t_0)\rightarrow f(t_0+\Delta t)=f(t_0)+f^\prime(t_0)\Delta t\rightarrow \cdots$.
Theoretically speaking, we expect that
this method can be applicable to the eigen structure ($|\varphi_n(\Lambda)\rangle$ and $\varepsilon_n(\Lambda)$)
of a large system with a good efficiency and a high fidelity.
However, the error accumulation through multiple adiabatic steps could limit the output fidelity
so that future experiments on large qubit devices would be necessary. % to explore the practical usage under good error correction.

\begin{acknowledgments}
The work reported here was supported by the National Key Research and Development
Program of China (Grant No. 2019YFA0308602, No. 2016YFA0301700), the National Natural
Science Foundation of China (Grants No. 12074336, No. 11934010, No. 11775129), the
Fundamental Research Funds for the Central Universities in China (2020XZZX002-01), and the Anhui Initiative
in Quantum Information Technologies (Grant No. AHY080000). Y.Y. acknowledge the
funding support from Tencent Corporation. This work was partially conducted at the University
of Science and Technology of China Center for Micro- and Nanoscale Research
and Fabrication.

\end{acknowledgments}

\begin{appendix}

\section{Single-Qubit Approximation}
\label{appA}

The adiabatic Hamiltonian of an $N$-qubit system can be in general expanded into
\be
H_\mathrm{ad}(t) = g^{(0)}(t) + \sum_{i=1}^N\sum_{j=1}^3 g^{(1)}_{ij}(t)\sigma_{i}^{j} + \sum_{i,k=1}^N\sum_{j,l=1}^3 g^{(2)}_{ij,kl}(t)\sigma_{i}^{j}\sigma_{k}^{l}+\cdots,
\label{eq_app01}
\ee
where $\{\sigma_{i}^{j}\!=\!X_i, Y_i, Z_i\}$ are the Pauli matrices acting on qubit $i$.
In our experiment, Eq.~(\ref{eq_app01}) is truncated up to the second order with nonzero coefficients in $\{g^{(0)}(t), g^{(1)}_{ij}(t), g^{(2)}_{ij,kl}(t)\}$.
To circumvent the instantaneous eigenstates in the counter-diabatic Hamiltonian, we apply a single-qubit approximation
to the multi-qubit interaction, e.g., $\sigma_{i}^{j}\sigma_{k}^{l}\approx \sigma^j_i +\sigma^l_k$~\cite{chenxiPRAPP20}.
The adiabatic Hamiltonian in Eq.~(\ref{eq_app01}) is decomposed into
$H_\mathrm{ad}(t)\!\rightarrow\!g^{(0)}(t)+\sum_{i=1}^N H_{\mathrm{ad}; i}(t)$ with $H_{\mathrm{ad};i}(t)=\vec{h}_i(t)\cdot \vec{\sigma}_i/2$.
%with $H_{\mathrm{ad}; i}(t)=\sum_{j=1}^3 [g^{(1)}_{i,j}(t)+\sum_{k,l}g^{(2)}_{i,j;k, l}(t)]\sigma^j_i$.
For conciseness, we introduce the vector of Pauli matrices, $\vec{\sigma}_i=X_i\vec{e}_1+Y_i\vec{e}_2+Z_i\vec{e}_3$, where $\{\vec{e}_{j=1,2,3}\}$
is the set of three unit vectors. The three components of the coefficient vector $\vec{h}_i(t)$ are given by $h_{i; j=1,2,3}(t)=g^{(1)}_{ij}(t)+\sum_{k,l}g^{(2)}_{ij,kl}(t)$.
For the $i$-th qubit, its counter-diabatic Hamiltonian is approximated as
\be
H_{\mathrm{cd};i}(t) \approx \frac{\vec{h}_i(t)\times\dot{\vec{h}}_i(t)}{2|\vec{h}_i(t)|^2}\cdot\vec{\sigma}_i.
\label{eq_app01a}
\ee
%which can be formally expanded into $H_{\mathrm{cd}; i}(t)=\sum_{j=1}^3 g_{\mathrm{cd};ij}(t)\sigma_{i}^{j}$.
The overall counter-diabatic Hamiltonian is approximated as
%\be
$H_\mathrm{cd}(t)\approx \sum_{i=1}^N\sum_{j=1}^3 g_{\mathrm{cd};ij}(t)\sigma_{i}^{j}$.
%\label{eq_app02}
%\ee
With the exact adiabatic and the approximate counter-diabatic parts, the total Hamiltonian for the subsequent digitized treatment is given by
\be
H_{\mathrm{tot}}(t)\approx g^{(0)}(t)+\sum_{i=1}^N\sum_{j=1}^3 [g^{(1)}_{i,j}(t)+g_{\mathrm{cd};i,j}(t)]\sigma_{i}^{j} + \sum_{i,k=1}^N\sum_{j,l=1}^3 g^{(2)}_{i,j;k, l}(t)\sigma_{i}^{j}\sigma_{k}^{l}.
\label{eq_app03}
\ee

\section{Bravyi-Kitaev Mapping}
\label{appB}

After the second quantization, the electronic Hamiltonian of the hydrogen molecule is formally written as
\be
H = \mathcal E_\mathrm{nucl}+ \sum_{i,j} h_{ij} a^+_i a_j +\sum_{i, j, k, l} h_{ijkl} a^+_i a^+_j a_k a_l,
\label{eq_app04}
\ee
where $h_{ij}$ and $h_{ijkl}$  are one- and two-electron integrals. Next we apply the Bravyi-Kitaev transformation~\cite{SeeleyJCP01},
where the fermion annihilation and creation operators are re-formulated as
\be
a_i   &=& \left[X_{U(i)}X_i Z_{P(i)}+iX_{U(i)}Y_i Z_{\rho(i)}\right]/2,  \\
a^+_i &=& \left[X_{U(i)}X_i Z_{P(i)}-iX_{U(i)} Y_i Z_{\rho(i)}\right]/2.
\ee
Here $U(i)$, $P(i)$ and $R(i)$ are the update, parity and remainder sets respectively and their explicit lists
can be found in Ref.~\cite{SeeleyJCP01}. The other set $\rho(i)$ is determined by
$\rho(i)=P(i)$ for $i\in \mathrm{even}$ and $\rho(i)=R(i)$ for $i \in \mathrm{odd}$. Through a straightforward derivation,
the Hamiltonian in Eq.~(\ref{eq_app04}) is simplified to
\be
H =g_0I+g_1Z_A+g_2Z_B+g_3Z_AZ_B+g_4X_AX_B+g_5Y_AY_B,
\label{eq_app05}
\ee
where the coefficients satisfy $g_1=g_2$ and $g_4=g_5$.
The extra condition of $|g_4|\gg|g_3|$ indicates that the three two-qubit couplings can be efficiently
simulated by $X_AX_B$ or $Y_AY_B$. In our experiment, we choose the latter two-qubit operator and the Hamiltonian is approximated as
\be
H = g_0  + g (Z_A +  Z_B) + g_{12} Y_AY_B,
\label{eq_app06}
\ee
The explicit dependence between $\{g_0, g, g_{12}\}$ and the internuclear distance $R$ in our experiment are cited from Ref.~\cite{CollessPRX18}.

\end{appendix}

%\end{CJK*}
\end{document}